\documentstyle[12pt]{article}
\author{Walter Simon \thanks{e-mail: simon@galileo.thp.univie.ac.at}\\
Institut f\"ur theoretische Physik der Universit\"at Wien\\
Boltzmanngasse 5\\A-1090 Wien}
\date{}
\title{Criteria for (in)finite extent of static perfect fluids}
\begin{document}

\maketitle

\begin{abstract}
In Newton's and in Einstein's theory we give criteria on the equation of
state of a  barotropic perfect fluid which guarantee that the corresponding 
one-parameter family of static, spherically symmetric solutions has finite extent. 
These criteria are closely related to ones which are known to ensure finite or
infinite extent of the fluid region if the assumption of spherical symmetry
is replaced by certain asymptotic falloff conditions on the solutions. 
We improve this result by relaxing the asymptotic assumptions. Our conditions on the equation of state 
are also related to (but less  restrictive than) ones under which it has been shown in Relativity that 
static, asymptotically flat fluid solutions are spherically symmetric. We present all these results in 
a unified way.

\end{abstract}

\newpage

\section{Introduction}

In this work we consider static, self-gravitating perfect fluids with a barotropic 
equation of state (EOS) $\rho(p)$ relating density and pressure, in Newton's theory (NT) and 
in General relativity (GR). We are thus dealing with time-independent,
globally regular solutions to the Euler-Poisson and the Euler-Einstein equations.
Some problems which have been studied in this context are the following:
\begin{description}
\item[A.] Do there exist (spherically symmetric, topologically $R^3 \times R$,
asymptotically flat) solutions with a given equation of state ?
\item[B.] Are all solutions (in particular all asymptotically flat ones) necessarily 
topologically $R^3 \times  R$ and spherically symmetric ?
\item[C.] Do fluid solutions (with parameters in some given range) have finite 
(or infinite) extent and a finite (or infinite) mass ?
\item[D.] Can solutions (if they exist) be (uniquely) determined by parameters
like surface potential, or mass (or central pressure or -density in the
spherically symmetric case)?
\item[E.] Are fluid solutions (with parameters in some given range) stable or unstable 
(against a certain classes of perturbations)?
\end{description}

Some of the questions listed above are clearly physically relevant when one develops models for 
stars or a model for the distribution of stars in a galaxy, while others are at least of 
mathematical interest. A recent review dealing with some of these issues can be found in \cite{BS}.
We will (after making a few general remarks), focus here on the question of finiteness (point C.), 
summarizing previous work as well as presenting new results. 

For smooth equations of state, existence of spherically symmetric solutions (point A) 
has been proven satisfactorily in NT and in GR \cite{RS}, while for discontinuous
equations of state an existence theory has been developed in NT only \cite{US}.
Regarding topology, partial results (in relativity) are given in \cite{AM}.
As to  spherical symmetry, a complete proof is available in NT \cite{LeLi}, 
but the problem is not really settled in GR (see \cite{LM2}, and below). 
Conjectures on points A and B under rather general conditions 
will be formulated in the next section.
In any case, it is useful to pose questions C, D and E in a general context, in particular without 
assuming spherical symmetry, since the latter is not required in some theorems.
We also note that the available results on spherical symmetry in GR not only have to be formulated 
somewhat differently for infinite and finite configurations, but the proof of the finite case 
actually goes the detour via a uniqueness theorem for given surface potential (point D). 
Thus, in GR points A and B have so far been studied jointly with (parts of) C and D.

Turning now to the details of the finiteness problem, an interesting aspect is the relevance of 
the behaviour of $\rho(p)$ for small $p$. In fact, by modifying the EOS suitably in just an arbitrarily small
neighbourhood of $p = 0$,  all corresponding static solutions become necessarily finite or infinite. 
To see this, recall that fluid balls with $\rho(0) \ne 0$ are always finite \cite{RS}, so we can simply 
set $\rho = \rho(\epsilon) = const.$ for $p \in [0,\epsilon]$ to ensure finiteness. On the other hand, (as we 
shall see below) we can always produce infinite solutions e.g. by gluing a piece with a linear EOS
$p = C.\rho$ for $p \in [0,\epsilon]$. 
We also recall that there are nontrivial criteria sufficient for finiteness of static fluid
configurations, for example (assuming spherical symmetry) the existence of the integral 
$\int_{0}^{p} dp'/\rho(p')^2$ for some $p > 0$ (due to Rendall and Schmidt \cite{RS}) or 
$(\rho/p) (dp/d\rho) =  \gamma + O(\rho^{\gamma - 1})$ for a constant $\gamma$
with $4/3 < \gamma < 2$ (due to Makino \cite{TM}), 
which only involve the behaviour of the EOS at small pressures. 

These facts nonwithstanding, it is important to realize that for the finiteness question the 
behaviour of $\rho(p)$ {\it for all $p$} is relevant in general. The striking examples 
in this context are the polytrope of index 5 in NT and the "Buchdahl" EOS in GR \cite{HB}, viz. 

\parbox{5.5cm}
{\begin{eqnarray} 
 p = \frac{1}{6}\rho_{-}^{-\frac{1}{5}} \rho^{\frac{6}{5}} \nonumber
\end{eqnarray}}
\hfill \parbox{7.5cm}
{\begin{eqnarray}
\label{Bucheos} 
 p = \frac{1}{6}(\rho_{-}^{\frac{1}{5}} - \rho^{\frac{1}{5}})^{-1}
\rho^{\frac{6}{5}}
\end{eqnarray}}\\
where $\rho_{-}$ is a positive constant, and $\rho < \rho_{-}$ in GR. All static solutions with these 
EOS have infinite extent but finite mass \cite{BS2}.
Criteria for (in)finiteness \cite{WS2} (discussed and generalized below) imply, in particular, 
that these solutions are "on the verge" of having (in)finite extent and (in)finite mass in the 
following sense: Suitable, but arbitrarily small modifications of the EOS performed in 
an arbitrarily small neighbourhood of an arbitrary value $p_{0}$ of $p$, necessarily force 
{\it all} solutions with  $p_{0} \in~ supp~\rho$  to be finite, while other small modifications 
near any  $p_{0}$  force {\it all} solutions  with $p_0 \in~ supp~\rho$ to have infinite mass. 

In this paper we discuss a theorem consisting of several parts. Under suitable conditions on the equation 
of state in each case, the theorem shows (roughly speaking) spherical symmetry from asymptotic flatness, 
finite and infinite extent of asymptotically flat fluid solutions, and finite extent in the spherically symmetric 
case. While the first part of the theorem is, in essence, just reproduced from \cite{BS1,LM1} and the second part is
a technically improved version of \cite{LM2,WS2}, the final part is the main new input of the present paper. 
The purpose of combining these three parts here is to exhibit the close relationship between the respective 
conditions on the equation of state (and in particular the role of (\ref{Bucheos}) as a limiting case). 
As a complement to the present work, Heinzle has recently obtained results on finiteness and infiniteness of 
static fluids \cite{MH}, assuming both asymptotic flatness and spherical symmetry, but less restrictive 
conditions on the equation of state than we do.

This report is organized as follows. In the next section we will give some preliminary material and 
quote the theorem discussed above. Crucial tools for the proof of the results on 
(in)finiteness are Pohozaev(-type) identities leading to virial(-type) theorems, which we discuss in Sect.3.
While in NT the virial theorem is an equality (without further assumptions), in GR one obtains in the same manner 
only  a "virial inequality", (but equalities can still be derived  when spherical symmetry is assumed \cite{MH}).
Correspondingly, the "general" Newtonian virial theorem yields criteria both for finiteness and infiniteness of the 
fluid configuration, whereas in GR one only obtains a finiteness criterium in the same fashion,
(but again an infiniteness criterium in the spherically symmetric case
\cite{MH}). However, even without this latter assumption a criterium 
for infiniteness  also exists in GR and is proven using the positive mass theorem with respect to a suitably 
conformally rescaled metric (a procedure which is, on the other hand, meaningless in NT). The proofs of the new 
parts of the main theorem are given in Sect.4. In the final section we expose our results in the light of the 
"quasipolytropic" family of EOS.

\section{The main theorem}

We will treat NT and GR in a "parallel" fashion as far as possible. This means that, if equations 
in NT and GR are direct analogues, we give them in left and right columns, respectively, and use 
the same symbols and the same numbers for corresponding quantities to facilitate comparison. 
In either situation, we consider a Riemannian manifold $({\cal M},g)$ with a scalar function 
$V$. In the Conjecture and the Theorem below, we will take $V$ and $g$ to be in some Sobolev 
space $L^q_{k+2}$ (using this symbol  we follow Lee and Parker \cite{LP}), with exponent 
$q\ge 4$ and with $k$, the degree of (weak) differentiability, equal to either $0,1$ or $2$. 
Furthermore, we consider on $({\cal M},g)$ the non-negative functions 
$p \in L^q_{k+1}$, and $\rho \in L^q_{k}$, related by an EOS $\rho(p)$ which is $C^{k-1}$
(if $k \ge 1$) and piecewise $C^k$. This means that for $k = 0$ we allow the density to "jump", 
while $k \ge 1$ forbids such jumps. We also require that $(g,V)$ satisfy on ${\cal M}$ 
(at least in a weak sense) the Euler-Poisson resp. the Euler-Einstein systems which we write as follows

\parbox{3.0cm}
{\begin{eqnarray}
\Delta V^{-1} &  = &  - 4\pi\rho \nonumber \\
R_{ij} & = & 0 \nonumber \\
D_{i} p & = &  \rho D_{i} V^{-1} \nonumber
\end{eqnarray}}
\hfill \parbox{9.5cm}
{\begin{eqnarray}
\label{DelV}
\Delta V & = & 4\pi (\rho + 3p) V \\
\label{Ricc}
R_{ij} & = & V^{-1} D_{i} D_{j} V + 4 \pi(\rho - p) g_{ij} \\
\label{Eu}
D_{i}p & = & - V^{-1} (\rho + p) D_{i}V.
\end{eqnarray}}

The reason why we use here $V^{-1}$ on the l.h. side will become clear shortly. 
In  the GR case, ${\cal M}$ can be understood as a hypersurface orthogonal to a timelike Killing 
vector $\xi$, and $V$ denotes the norm of $\xi$. The covariant derivative $D_{i}$, 
the Laplacian  $\Delta = D_{i}D^{i}$ and the Ricci tensor $R_{ij}$ refer to $g$. As to the Euler 
equation, there is a difference between NT and GR: 
While in the former case we require that $p \in L^{q}_{1}$ and postulate (\ref{Eu}), 
the corresponding (weak) version in GR can be derived as follows.\\ \\
{\it Lemma 1.} (C.f. Remark 1.8. of \cite{AM}.) 
Let $({\cal M},g,V)$ be a $L^q_{2}$ solution which satisfies the
GR-side of the field equations (\ref{DelV}) and (\ref{Ricc}) weakly. 
Then $p \in L^q_{1}$, and Euler's equation (\ref{Eu}) also holds weakly.\\ \\
{\it Proof.} 
Since $C^{\infty}_0$ functions lie dense in $L^q_{k}$ we can approximate $g$
by a sequence of such functions. For any test function $\phi \in C_{0}^{\infty}$ we
have then the following weak form of the Bianchi identity

\begin{equation}
\label{Bi}
\int_{\cal  M}D^{i}\phi^{j}(R_{ij} - \frac{1}{2} R g_{ij}) d{\cal V} = 0
\end{equation}

where $d{\cal V}$ is the volume element w.r. to $g$. 
Approximating now also $V,p$ and $\rho$ by $C_{0}^{\infty}$-functions
we easily obtain the required result.   $\hfill \Box$ \\ 

While we allow $\rho$ to have zeros in general (in particular at $p=0$) we always assume 
that the integrals 

\parbox{5.5cm}
{\begin{eqnarray}
 \Gamma & =  & \int_{0}^{p}\frac{dp^{\prime}}{\rho(p^{\prime})} 
\nonumber
\end{eqnarray}}
\hfill
\parbox{7.5cm}
{\begin{eqnarray}
\label{Gam}
\Gamma & = & \int_{0}^{p}\frac{dp^{\prime}}{\rho(p^{\prime}) + p^{\prime}}
\end{eqnarray}} \\
exist for some interval $p \in [0,p_{max}]$. (The integral does not exist e.g. for linear equations 
of state $\rho = C.p$).
Obviously  $\Gamma$ is a continuous function of $p$ and also on ${\cal M}$. In NT, from Euler's 
equation, we have $D_{i}(V^{-1}- \Gamma) = 0$ and so the expression in parenthesis is a constant
$V_s^{-1}$ equal to $V^{-1}$ at $p = 0$, whence $V_s$ is called the surface potential. In GR, the same 
conclusions can be drawn from $D_{i}(V e^{\Gamma}) = 0$. Integrating (\ref{Eu}) we obtain

\parbox{5.5cm}
{\begin{eqnarray}
 \Gamma = V^{-1} - V_{s}^{-1}
\nonumber
\end{eqnarray}}
\hfill
\parbox{7.5cm}
{\begin{eqnarray}
\label{Euint}
\Gamma = \mbox{ln} V_s - \mbox{ln} V.
\end{eqnarray}} \\
$V$ is related to the standard Newtonian potential $U$ by $U = 1 - V^{-1}$. 
To formulate our theorems we now introduce some quantities defined from the EOS, and proceed 
with a lemma on their mutual relationship. If $\rho(p)$ is $C^0$ and piecewise $C^1$
we define

\parbox{5.5cm}
{\begin{eqnarray} 
\kappa  =  \frac{d\rho}{dp} \nonumber
\end{eqnarray}}
\hfill \parbox{7.5cm}
{\begin{eqnarray} 
\label{kap}
\kappa  =  \frac{\rho + p}{\rho + 3p} \frac{d\rho}{dp}
\end{eqnarray}}\\
and if $\rho(p)$ is $C^{k-1}$ (for $k = 1,2$) and piecewise $C^k$ (for $k = 0,1,2$),$\hfill {}$

\parbox{5cm}
{\begin{eqnarray} 
I_{0} & = & \rho \Gamma -  6 p \nonumber\\
I_{1} & = &  6 p \kappa  - 5 \rho \nonumber\\
I_{2} & = &   5 \rho \frac{d\kappa}{dp} + \kappa^2 \nonumber
\end{eqnarray}}
\hfill 
\parbox{8cm}
{\begin{eqnarray} 
\label{I0}
I_{0} & = & \rho (e^{\Gamma} - 1) - 6 p \\
\label{I1}
I_{1} & = &  6 \frac{\rho + 3p}{\rho} p \kappa - 5 \rho  \\
\label{I2}
I_{2} & = & 5 (\rho + p)\frac{d\kappa}{dp} + \kappa^2 + 10 \kappa.
\end{eqnarray}}

The static spherically symmetric solutions corresponding to $I_0 \equiv 0$ and $I_1 \equiv 0$ 
are known explicitly and read (c.f. \cite{SC} for NT and \cite{BS2,HB} for GR),

\parbox{5.5cm}
{\begin{eqnarray} 
1 - V^{-1} =  - \frac{M}{\sqrt{\frac{4\pi}{3}\rho_{-}M^4 + r^2}}\nonumber
\end{eqnarray}}
\hfill \parbox{7.5cm}
{\begin{eqnarray}
\label{Buchsol} 
1 - V = \frac{M}{\sqrt{\frac{4\pi}{3}\rho_{-}M^4 + r^2} + \frac{M}{2}},
\end{eqnarray}}\\
where $M$ is the mass which can take any positive value in NT. 
In GR, $M$ is bounded from {\it below} by $3M^{-2} < 16\pi
\rho_{-}$, and the metric on ${\cal M}$ is $g = [(2/(1+V)]^4 \delta$ where 
$\delta$ is the Kronecker symbol.

We continue with a result on the relation between the quantities $I_k$. \\ \\
{\it Proposition 1.} 
Assume that $\rho > 0$ for $p > 0$, and that $\lim_{p \rightarrow 0} \rho^{-1} p $ exists.
Let  $\{I_{k} \ge 0\}$, $\{I_{k} \equiv 0 \}$ and $\{I_{k} \le 0\}$
denote the sets of equations of state which are $C^{k-1}$ (for $k = 1,2$), piecewise $C^{k}$
for $k = 0,1,2$  and which satisfy, for all $p \in [0,p_{max}]$, $I_{k} \ge 0$,  $I_{k} \equiv 0$, or $I_{k} \le 0$,
respectively.  Then

\begin{eqnarray}
\label{Ige}
{} & {} & \{I_{1} \ge 0 \}  \subseteq  \{I_{0} \ge 0 \}\\
\label{Ieq}
\{I_{2} \equiv 0 \} & \supseteq & \{I_{1} \equiv 0 \} \equiv  \{I_{0} \equiv 0 \}\\
\label{Ile}
\{I_{2} \le 0 \} & \subseteq & \{I_{1} \le 0 \}  \subseteq  \{I_{0} \le 0 \}.
\end{eqnarray}\\
{\it Proof.} These results are due to the fact that the $I_{j}$ are in a certain sense the integrals of 
$I_{j + 1}$. In particular, from (\ref{Eu}), (\ref{Euint}), (\ref{I0}) and (\ref{I1}) we obtain

\parbox{5.5cm}
{\begin{eqnarray} 
\rho \frac{d}{dp} [\rho^{-1} I_0] = \rho^{-1} I_1 \nonumber
\end{eqnarray}}
\hfill \parbox{7.5cm}
{\begin{eqnarray}
(\rho + p) e^{\Gamma} \frac{d}{dp} [\rho^{-1} e^{-\Gamma} I_0] = \rho^{-1} I_1.
\end{eqnarray}} \\
To show (\ref{Ige}) and the second inclusion in (\ref{Ile}) we note that $\rho^{-1}I_0$ vanishes at 
$p = 0$, which is obvious when $\rho_s \ne 0$. As to the case $\rho_s = 0$, we first note that the existence 
of $\Gamma$ and of $\lim_{p \rightarrow 0} \rho^{-1}p $ implies $\lim_{p \rightarrow 0} \rho^{-1}p = 0$ 
(c.f. \cite{MH}, Appendix A) and so the assertion follows from (\ref{I0}).
Therefore, when $I_{1} \ge  0$, $I_{1} = 0$ or $I_{1} \le 0$, the same holds for $I_0$.

Each of the conditions $I_{1} \equiv 0$ and $I_{0} \equiv 0 $ (together with the differentiability 
assumptions) characterizes the 1-parameter families of EOS (\ref{Bucheos}) and so the sets 
 $ \{I_{1} \equiv 0 \} \equiv  \{I_{0} \equiv 0 \}$ of corresponding
solutions are given by (\ref{Buchsol}). The condition $I_2 = 0$ admits the case $\rho = const.$ 
and a two-parameter family of EOS which reads $p = 1/6(\rho_{-}^{-1/5}\rho^{6/5}
- \rho_{+})$ in NT (where $\rho_{-}$ and $\rho_{+}$ are constants with $\rho_{-} >\rho_{+}$), 
while in GR it is given in \cite{WS1} (together with the corresponding static solutions).

Finally, the first inclusion in (\ref{Ile}) is proven in \cite{BS1} (in a
similar way as above).
$\hfill \Box$ \\

We now impose falloff conditions on $g$ and $V$ in terms of weighted Sobolev spaces
$L^{q}_{k,\beta}$ (defined as in Def. (2.1) of Bartnik  \cite{RB}, and Sect. 9 of Lee and 
Parker, \cite{LP} following again the notation of the latter paper). 
\\ \\
{\it Definition 1.}  
A $L^{q}_{2}$ solution $({\cal M},g,V)$ of (\ref{DelV}), (\ref{Ricc}) is called asymptotically flat (AF) if 
\begin{enumerate}
\item The "end" ${\cal M}^{\infty} = {\cal M} \setminus \{\mbox{a compact set} \}$ is 
diffeomorphic to $R^3 \setminus {\cal B}$ where ${\cal B}$ is a closed ball. 
\item On ${\cal M}^{\infty}$, $1-V = o(1)$, and the metric $g$ satisfies, for some $q \ge 4$ 
and $\alpha > 0$,
\begin{equation}
\label{gaf}
g - \delta  \in  L^{q}_{2,\alpha}. 
\end{equation}
\end{enumerate}
Thus, in this definition of AF we have supplemented Bartnik's falloff condition \cite{RB} for $g$
by the requirement that $V \rightarrow const.$ at infinity (with the convention that
$const. = 1$).\\ \\
{\it Definition 2.}  An AF solution $({\cal M},g,V)$ of (\ref{DelV}), (\ref{Ricc}) 
is called asymptotically flat with mass decay conditions (AFMD) if  on ${\cal M}^{\infty}$ 
the fluid variables satisfy, for some $q \ge 4$ and $\alpha > 0$,
\begin{eqnarray}
\label{rhoac}
\rho & \in & L^q_{0,-3 -\alpha}, \\ 
\label{pac}
p & \in  & L^q_{0,-3-\alpha}.
\end{eqnarray}
Here the name "mass decay conditions" for these falloff conditions is inspired by
similar requirements on the Ricci scalar in eqns.(4.4) of \cite{RB}.

We remark that, to obtain our results in NT, it would be sufficient to impose (\ref{rhoac}) 
and (\ref{pac}) with $"3"$ replaced by $"5/2"$, which would not a priori restrict us to finite 
mass. In order not to spoil the "parallel" presentation of NT and GR, we just restrict ourselves 
to remarks on this option (after the formulation of the theorem).

Together with the field equations (\ref{DelV}), (\ref{Ricc}), the AFMD
conditions lead to the following stronger falloff properties. (We only give the
formulation in GR explicitly; the Newtonian case arises via obvious
simplifications.) \\ \\
{\it Lemma 2.}
Let $({\cal  M},g,V)$ be an AFMD GR solution. 
Then there exist a harmonic coordinate chart $({\cal M}^{\infty},g)$ and a constant 
$M$ ("the mass") such that 

\begin{eqnarray}
\label{Vas}
1 - V - \frac{M}{r} & \in & L^{q}_{2,-1-\alpha} \\
\label{gas}
g - (1 + 2M/r)\delta & \in & L^{q}_{2,-1-\alpha}\\
\label{pas}
p & \in & L^{q}_{1,-4-\alpha}
\end{eqnarray} \\
{\it Proof.} Inserting $1 - V =  o(1)$,  (\ref{rhoac}) and (\ref{pac}) in (\ref{DelV}) 
we have  $\Delta V \in L^q_{0,-3 -\alpha}$. 
Inverting the Laplacian with Proposition 2.2 of \cite{RB} we obtain
(\ref{Vas}). The next step is to insert (\ref{rhoac}), (\ref{pac}) and (\ref{Vas}) in (\ref{Ricc}) 
which yields $R_{ij} \in L^q_{0,-3-\alpha}$. By Proposition  3.3 of \cite{RB} we can now pass to 
harmonic chart on ${\cal M}^{\infty}$ and get (\ref{gas}). 
Finally, (\ref{pas}) is obvious after inserting (\ref{Vas}) and (\ref{gas}) in (\ref{Eu}). 
$\hfill \Box$ \\ 

If we replace "3" by "5/2" in the definition  of AFMD, we get Lemma 2 without the mass term, 
and with falloff $L^{q}_{2,-1/2-\alpha}$ for $V,g$ and $L^{q}_{1,-3-\alpha}$ for $p$.

We now formulate a conjecture on points A,B and C raised in the
introduction, and compare it with our theorem. \\ \\
{\it Conjecture.}
Let $\rho(p)$ be a piecewise $C^0$ equation of state with $\rho \ge 0$, $p \ge 0$. Then 
\begin{description}

\item[A:] There exists a 1-parameter family of spherically symmetric, AF 
$L^{q}_{2}$ solutions $({\cal M},g,V)$ of 
(\ref{DelV},\ref{Ricc},\ref{Eu}).

\item[B:] Every $L^q_{2}$-solution $({\cal M},g,V)$ is necessarily spherically
symmetric.

\item[C:]{\bf I}. If $I_{0} \le 0$  holds for $p \in [0, p_{max}]$,
then the solution either has finite extent or $\rho(p)$ satisfies 
$I_{0} \equiv 0$. 
\item[{}~~~~{}II.]
If $I_{0} \ge 0$  holds for $p \in [0, p_{max}]$,
then either $\rho(p)$ satisfies $I_{0} \equiv 0$ or the solution is not AF.

\end{description}

As remarked in Sect. 1, part A was shown under slightly stronger assumptions in NT \cite{US} and for smooth EOS in GR
\cite{RS}. We refer to these papers for details. As to spherical symmetry (B), the
following theorem requires
AFMD as well as a condition on the EOS, while the results on finiteness require either AFMD or spherical symmetry,
together with stronger assumptions on the EOS than conjectured above.\\ \\
{\it Theorem.} Let $\rho(p)$ be given with $p \ge 0$ and $\rho \ge 0$
and assume that there is a solution $({\cal M},g,V)$ of (\ref{DelV},\ref{Ricc},\ref{Eu}) specified below.

\begin{description}

\item[B:]
Assume that $\rho(p)$ is $C^1$ with $d\rho/dp > 0$ and piecewise $C^{2}$
with $I_2 \le 0$  for all $p \in [0, p_{max}]$, and assume that the solution is  $L^{q}_4$ ($q \ge 4$) and AFMD.

Then the solution is spherically symmetric. 

\item[C:]{\bf I.} If one of the following conditions 1. or 2. holds 

\begin{description}
\item [1.] $\rho(p)$ is piecewise $C^0$, and $I_{0} \le 0$ holds for $p \in [0, p_{max}]$.
The solution is AFMD. 
\item[2.]
$\rho(p)$ is $C^0$, piecewise $C^1$ with $d\rho/dp > 0$, and $I_{1} \le 0$ holds for $p \in [0, p_{max}]$.
The solution is  $L^q_{3}$ ($q \ge 4$) and spherically symmetric.
\end{description}
Then the solution either has finite extent, or $\rho(p)$ satisfies $I_{0} \equiv 0 $.\\
\item[{}~~~~{}II.] Assume that $\rho(p)$ is piecewise $C^0$, and $I_{0} \ge 0$ 
holds for all $p \in [0, p_{max}]$.

Then either $I_0 \equiv 0$,  or the solution is not AFMD.

\end{description}

Some comments on this theorem are in order.

Whenever $I_{0} \equiv 0$, then the EOS reads (\ref{Bucheos}), 
and the solutions are given by (\ref{Buchsol}).

Part B. of this theorem has been proven in \cite{LM2} in the smooth case, but extends straightforwardly 
to the present setting. 

Part C.I.1. has in essence been shown in \cite{WS2}, while C.II is proven in \cite{WS2} and 
(assuming $I_{1} \ge 0$) in \cite{LM1}. 
Compared to these papers, we set out here from weaker assumptions on differentiability and on the falloff
(as to NT, c.f. the  remarks after Definition 2 and after Lemma 2). 

The main new result of the present paper  is part C.I.2. 
Its proof goes, in essence, the detour via C.I.1, i.e. we  first show that assumptions C.I.2 
(together with the field equations) imply falloff conditions similar to (\ref{Vas}), (\ref{gas}) and 
(\ref{pas}) which can be used alternatively to the AFMD ones in C.I.1. in the spherically symmetric case. 
Since we also know from Proposition 1 that  $I_{1} \le 0$ implies $I_{0} \le 0$, finite extent 
of the solution follows. 

We note that  C.I.1 and C.I.2 are also related in the sense that they require similar auxiliary results, 
(in particular Lemma 4) and so we found it useful to repeat (from \cite{WS2}) the whole proof of C.I.1 here. 

A natural extension of part C of the above theorem has been obtained by Heinzle \cite{MH}.
From the equation of state he defines quantities $J_{-1}$ in NT and $J_{0}$ in GR whose signs again 
determine the (in)finiteness properties of solutions which are now assumed to be both spherically symmetric 
and asymptotically flat. Moreover, as an extension of Proposition 1, it is shown in \cite{MH} that if $I_0$ 
has a sign (and if $\lim_{p \rightarrow 0} \rho^{-1}p$ exists), then  $J_{-1}$ and $J_{0}$ 
have the same sign as $I_0$. Hence these quantities extend our "$I_k$ series" in a natural manner, and
Heinzle's (in)finiteness results require rather weak conditions on the equations of state.  
 
\section{The virial theorem}

This section contains two easy technical lemmas followed by the "virial theorem" (as
a proposition). The latter is a tool for the proof of the Theorem in Sect.4.\\ \\
{\it Lemma 3.} (The modified Pohozaev identity \cite{WS2}; compare \cite{SP}).\\
On $({\cal M},g)$ with ${\cal M} \sim R^3$ and $g$ flat, let $\xi_{i}$ be a dilation, i.e. 
$D_{(i}\xi_{j)}=\delta_{ij}$ (in Cartesian coordinates $x^{i}$, $\xi_{i} = x^{i}$). 
Assume that $\sigma = \sigma(X)$ with $X \in L^q_2$, $\sigma \in L^q_0$ $(q \ge 4)$ satisfies 
$\Delta X = 4\pi \sigma$, and define functions $\tau(X) = \int^{X} \sigma(X') dX'$ and $Z = D_{j}X
D^{j}X$. (Here $\tau$ is defined only up to an additive constant). Then we have

\begin{equation}
\label{Poh}
D_{i}[(\xi^{j}D_{j}X + \frac{1}{2}X)D^{i}X -
\frac{1}{2} Z \xi^{i} + 4\pi \tau \xi^{i}] = 2\pi(\sigma X + 6\tau).
\end{equation}\\
This  Lemma, which is proven by straightforward computation, will be applied below in NT, with 
$X = 1- V^{-1}$, $\sigma = \rho$ and $\tau = p$. 

In spaces in which there exists a (general) conformal Killing vector $\xi$, 
(i.e. ${\cal C}[D_{i}\xi_{j}] = 0$ where ${\cal C}$ denotes the symmetric, trace-free
part) there is an obvious generalization of (\ref{Poh}). 
A more subtle generalization is the following.\\ \\
{\it Lemma 4.} 
Let $({\cal M},g')$ be a Riemannian 3-manifold with $g' \in L^q_2$ and Ricci-tensor
and -scalar $R_{ij}^{\prime}$, $R^{\prime}= g^{\prime ij} R_{ij}^{\prime}$ and 
$B_{ij}^{\prime}={\cal C}^{\prime}[R_{ij}^{\prime}]$
(where ${\cal C}^{\prime}$ denotes the symmetric,trace-free part w.r. to $g'$).
Assume further that there exist a vector field $\eta^{\prime i} \in L^{q}_{1}$ and a function
$\alpha' \in L^q_{0}$ such that 
\begin{equation}
\label{Deta}
{\cal C}^{\prime}[D_{i}^{\prime}\eta_{j}^{\prime}] = 
\alpha^{\prime}B_{ij}^{\prime}.
\end{equation}

Then
\begin{equation}
\label{DR}
D^{\prime i}[(R_{ij}^{\prime} -
\frac{1}{2}g_{ij}^{\prime}R^{\prime})\eta^{\prime j}] 
 =  \alpha^{\prime}B_{ij}^{\prime}B^{ij\prime} -
\frac{1}{6}R^{\prime}D_{i}^{\prime}\eta^{\prime i} 
\end{equation}

and if $g' \in L^q_{3}$,
\begin{equation}  
\label{DB}
D^{\prime i}[B_{ij}^{\prime} \eta^{\prime j}]   = 
 \alpha^{\prime}B_{ij}^{\prime}B^{ij\prime} +
\frac{1}{6}\eta^{\prime i}D_{i}^{\prime} R^{\prime}. 
\end{equation}

The derivatives on the l.h. sides on (\ref{DR}) and (\ref{DB}) are to be understood in the 
weak sense. \\ \\
{\it Proof.} (\ref{DR}) and (\ref{DB}) follow from (\ref{Deta}) by approximating $\eta'$ with a 
sequence of $C^{\infty}_0$ functions, and from the weak Bianchi identity (\ref{Bi}) with respect 
to $g'$.  $\hfill \Box$ \\ 

If there exists a conformal Killing vector $\xi$ then we could in principle take $\eta = \xi$
in Lemma 4. Instead, we will make later use of this lemma with the more sophisticated choice

\parbox{5.5cm}
{\begin{eqnarray}
g' & = & (1 - V^{-1})^4 g \nonumber \\[0.4cm] 
\eta_{i}' & = & D_{i}(1 - V^{-1})^2 \nonumber\\[0.4cm] 
\alpha' & = & - (1 - V^{-1})^2 \nonumber
\end{eqnarray}}
\hfill 
\parbox{7.5cm}
{\begin{eqnarray} 
\label{gpr}
g' & = & \frac{1}{16}(1 + V)^4 g \\
\label{etpr}
\eta_{i}' & = & \frac{1 + V}{(1 - V)^3} D_{i}V \\ 
\label{alpr}
\alpha' & = & \frac{V(1 + V)^2}{(1 - V)^4}.
\end{eqnarray}}

As to the GR case of (\ref{DR}) we note that, using (\ref{gpr})-(\ref{alpr}),
 the l.h. side can be rewritten (after multiplying with $(det~ g'/det~ g)^{1/2}$) as 

\begin{equation}
\label{DW}
D_{i} \left[ \frac{1}{1 - V^2)^{2}} \left( V^{-1} D^{i}W + 8 \frac{ W D^{i}V}{1 - V^2} -
8\pi \left(\rho + \frac{1 - 5V}{1-V}p \right) D^{i} V  \right) \right],
\end{equation}\\
where $W = D_i V D^i V$. The quantities on the r.h. side of (\ref{DR}) read explicitly

\begin{eqnarray}
\label{Rp}
R^{\prime} & = & 256\pi (1 +  V)^{-5} [\rho (1 -  V) - 6pV]\\
\label{Detap}
D^{\prime}_{i}\eta^{\prime i} & = &
32 (1 - V^2)^{-4} [ 2\pi V(1 - V^{2})(\rho + 3p) + 3W],
\end{eqnarray}\\
and so $D^{\prime}_{i}\eta^{\prime i} \ge 0$.

We can now show the following \\ \\
 {\it Proposition 2.} (The "Virial theorem")\\
Let $({\cal M},g,V)$ be AFMD. Then there is a $B \ge 0$ such that

\parbox{5.5cm}
{\begin{eqnarray}
\int_{\cal M} [\rho (1 - V^{-1}) + 6p] d{\cal V} = 0  \nonumber
\end{eqnarray}}
\hfill 
\parbox{7.5cm}
{\begin{eqnarray} 
\label{vir1}
\int_{\cal M} B [\rho (1 - V^{-1}) + 6p] d{\cal V} \le 0. \end{eqnarray}}\\ 
{\it Proof.} In NT we integrate the Pohozaev identity (\ref{Poh}), apply the divergence 
theorem and note that the resulting surface terms vanish by virtue of the asymptotic 
properties derived in Lemma 2 (see also the remark after this lemma).
In GR we apply the same reasoning to (\ref{DR}), using (\ref{Rp}),
$B = (128 V \pi/3)(1 + V)^{-5} D_{i}'\eta^{\prime i}$ and 
again Lemma 2 for the surface terms.        $\hfill \Box$

\section{Proof of the main theorem}

We first sketch the ideas of the proof. For part C.I.1 in general,  and also for part
C.II in NT we can use the virial theorem described above (which requires asymptotic flatness). 
We find that the sign of  $I_0$, together with $V_s < 1$ or $V_s = 1$, determine the
signs of the r.h. sides of (\ref{vir1}), which directly leads to the required conclusions. 
To show C.I.2 we introduce the quantities

\parbox{4.5cm}
{\begin{eqnarray}
\widehat W & = & (1 - V^{-1})^{-4} W \nonumber \\
\widehat g & = & (1 - V^{-1})^4 g \nonumber \\
\widehat W_{0} & = & \frac{4\pi}{3} \int_{V_{c}}^{V} \frac{\rho V^4}{(1 - V)^4}dV \nonumber
\end{eqnarray}}
\hfill 
\parbox{8.5cm}
{\begin{eqnarray}
\label{What}
\widehat W & = & (1 - V^2)^{-4} W\\
\label{ghat}
\widehat g & = & V^{-2} (1 - V^{2})^{4} g \\
\label{W0hat}
\widehat W_{0} & = & \frac{4\pi}{3} \int_{V_{c}}^{V} \frac{V(\rho + 3p)}{(1 - V^2)^4}dV,
\end{eqnarray}}\\
where $V_c$ is a constant specified below.
We then employ (\ref{DB}), again with the choices (\ref{gpr}),(\ref{etpr}) and (\ref{alpr}). 
As the r.h. side now contains $D_i' R'$ (instead of $R'$ in (\ref{DR})), controlling the sign now requires, 
(instead of $I_0 \le 0$), the more restrictive condition $I_1 \le 0$. In fact, assuming now 
also that the fluid region extends to infinity (i.e. $V_s = 1$) we can write (\ref{DB}) in 
the form 

\begin{equation}
\label{DelW}
\widehat \Delta (\widehat W - \widehat W_{0})  \ge 0
\end{equation} \\
with $\widehat W_0 \ge C^2 = const. > 0$. 
If we were to assume that $({\cal M},g,V)$ is AFMD, we could now form a compactification 
$\widehat {\cal M} = {\cal M} \cup \Lambda$ of this manifold by adding the point 
at infinity $\Lambda$, and  $\widehat W$ , $\widehat W_0$ and the metric $\widehat g$ would have 
$C^1$-extensions to $\Lambda$. Then the maximum principle would immediately lead again to the 
conclusions of C.I.1. Without asymptotic conditions we can still apply the maximum principle to 
(\ref{DelW}) on a finite domain ${\cal D}$. Assuming also spherical symmetry and taking
$V_c$ in (\ref{W0hat}) to be the value of $V$ at the centre, we conclude that 
$\widehat W\ge \widehat W_0 > C^2$, which gives $1 - V \le 1/C.r$ on a ball ${\cal B}_r$ of radius $r$. 
Thus, extending ${\cal B}_r$ we control the asymptotic behaviour of $V$. With some 
further technical manipulations based on the fact that  $\Delta V$ is positive, we can then prove that 
$({\cal M},g,V)$ satisfies asymptotic conditions similar to the  AFMD ones. 
These in fact guarantee the existence of a $C^1$ compactification as well as the vanishing of the 
surface terms in the integrals of (\ref{Poh}) and (\ref{DW}). We then obtain the required conclusion 
$I_1 \equiv 0$ either from the maximum principle or by applying the reasoning of C.I.1.

We remark that, from (\ref{DW}), we could also write (\ref{DB}) in the form 
$\widehat \Delta (\widehat W - \widehat W_1)  \ge 0$ for some $\widehat W_{1}(V)$
(assuming infinite extent and  $I_1 \le 0$).
This relation can, however, not replace (\ref{DelW}) in the proof sketched above as we do not have 
control over the sign of $\widehat W_1$ (whereas $\widehat W_0  > 0$). 

Finally, to show part C.II in the Einstein case, we use the rigidity case of the positive mass theorem 
on $({\cal M},g')$.

We now turn to the details.\\ \\
{\it Proof of part C.I.1 of the Theorem}.
Using (\ref{Euint}),(\ref{I0}) and the virial theorem (\ref{vir1}), we find that

\parbox{5.5cm}
{\begin{eqnarray} 
\int_{\cal M} [\rho (1 - V_s^{-1}) - I_{0}] d{\cal V}  =  0 \nonumber 
\end{eqnarray}}
\hfill 
\parbox{7.5cm}
{\begin{eqnarray}
\label{vir2} 
\int_{\cal M}B [(V_s - 1)(\rho + 6p) -  I_{0}] d{\cal V} \le 0,
\end{eqnarray}} \\
where $I_{0}$ has here to be considered as a function on ${\cal M}$.
The statement of the theorem now follows immediately from (\ref{vir2}).\\ \\
{\it Proof of  part C.II}.
In NT the second part of the theorem also follows easily from
(\ref{vir2}). As to the GR case, we find from (\ref{Vas}) that the metric 
$g'$ is $L^q_{2,\alpha}$, has vanishing mass, and $R' \geq 0$ due to (\ref{Euint}), 
(\ref{I0}), (\ref{Rp}) and $I_{0} \geq 0$. 
The "vanishing mass-" case of the positive mass theorem \cite{SY} hence implies that 
$g'$ is flat, and so $R'$ vanishes identically. 
Using (\ref{Rp}) together with (\ref{Eu}) gives the Buchdahl EOS.\\ \\
{\it Proof of part C.I.2.}
We show that the requirements imply asymptotic conditions which in essence reduce the problem to 
C.I.1. We treat NT and GR simultaneously.  

Observing from (\ref{Rp}) that the Ricci scalar $ R'$ with respect to $g'$ is a
function of $V$, we first show that $I_{1} \le 0$ implies a lower bound for $dR'/dV$. 
To do so, we note first that the existence of $\Gamma$ and $d\rho/dp > 0$
imply that $\lim_{p\rightarrow 0} \rho^{-1} p = 0$ (c.f. Appendix A of \cite{MH}).
By a straightforward computation, using $I_{1} \le 0$, $\rho^{-1}I_0 \le 0$ 
(which follows from Proposition 1 and form the previous remark), and (\ref{Euint})
we get

\parbox{5.5cm}
{\begin{eqnarray}
\lefteqn {\frac{(1 - V)^6}{32 \pi V^4 \rho \kappa} \frac{d}{dV}  R' = {}}
\nonumber\\
 & & {} = 5 \kappa^{-1} + 1 - V^{-1}  \ge {}
\nonumber\\
& & {} \ge \frac{6p}{\rho} + 1 - V^{-1} \ge {}
\nonumber\\
 & & {} \ge  1 - V^{-1} + \Gamma \nonumber\\
 & & {} \ge  1 - V_s^{-1} 
\nonumber
\end{eqnarray}}
\hfill
\parbox{7.5cm}
{\begin{eqnarray}
\label{dVRhat}
\lefteqn{ \frac{V(1 + V)^6}{256 \pi (\rho + 3p)\kappa} \frac{d}{dV} R' = {}}
\nonumber\\
 & & {} = V^2 (10 \kappa^{-1} + 1) - 1 \ge {}
\nonumber\\
& & {} \ge 12 V^2 \left[ 3 \frac{p^2}{\rho^2} + \frac{p}{\rho} + 1 \right] - 1 \ge {}
\nonumber\\
 & & {} \ge  V^2  e^{2\Gamma} - 1 \nonumber\\
 & & {} \ge  V_s^{2} - 1.
\end{eqnarray}}

We now assume that the fluid extends to infinity, (i.e. $V_{s} = 1$), 
which will lead to a contradiction. It follows from (\ref{dVRhat}) that $dR'/dV \ge 0$.
Restricting ourselves to spherical symmetry we have $\widehat W_{0} = 0$ at the centre by
definition, and $\widehat W =  0$ since $V$ takes its minimum there. Applying the maximum principle 
to (\ref{DelW}) on any ball ${\cal B}_r$ (bounded by $r$  = const.) we obtain 
$\widehat W \ge \widehat W_{0}$ on ${\cal B}_r$. Next, since the integrand (\ref{W0hat}) is non-negative, 
$W_{0}(V)$ increases monotonically with $V$ (and with $r$); in particular there is a constant $C > 0$ 
such that $\widehat W \ge \widehat W_{0} \ge C^2$.

Below we give the proof of the relativistic case explicitly. (The Newtonian case can easily be obtained 
by suitable simplifications).

We introduce coordinates such that
\begin{equation}
\label{gsph}
ds^2 = g_{ij}dx^i dx^j =  (1 - 2 r^{-1} m(r))^{-1}dr^2 + 
r^2 (d\theta^2 + sin^2 \theta d\phi^2)
\end{equation}
where $m(r) = 4\pi \int_{0}^{r} r^{\prime 2} \rho(r^{\prime}) dr^{\prime}$.
Writing out the bound $\widehat W \ge C^2$ explicitly we obtain, since 
$g^{rr} < 1$,

\begin{equation}
\label{dVbd}
\frac{d}{dr} \frac{1}{1 - V} =  \frac{1}{(1 - V)^{2}} \frac{dV}{dr} \ge 
\frac{\sqrt{g^{rr}}}{(1 -  V^{2})^2} \frac{dV}{dr} = \sqrt{\widehat{W}} \ge C
\end{equation} \\
which implies that 
\begin{equation}
\label{Vbd}
1 - V \le  \frac{1}{C.r}.
\end{equation}

The crucial step is now to get an upper bound for $r^2 dV/dr$.  
Writing (\ref{DelV}) as 
\begin{equation} 
\label{d2V} 
\frac{d}{dr}\left( r^2 \sqrt{g^{rr}} \frac{dV}{dr} \right)  = 4 \pi V (\rho + 3p)
r^2 \sqrt{g_{rr}}
\end{equation} \\
and noting that the r.h. side is non-negative for $p > 0$, we find 
 that $dV/dr$ is positive and that $r^2 \sqrt{W}$ is strictly 
monotonically increasing. (Note that, from  Theorem 2 of \cite{RS}, 
$g^{rr}$ does not go to zero for finite $r$. This argument also holds 
under the present differentiability assumptions). We now show by an 
indirect argument that $r^2 \sqrt{W}$ is bounded as $r \rightarrow \infty$. 
Assuming the contrary, namely that
\begin{equation}
\label{dVdr1}
\forall~D > 0~~\textrm{and}~~\forall~r_1~~\exists~~r_0 > r_1~~~
\textrm{such that}~~ r^2 \sqrt{W}|_{r_0} > D
\end{equation}
we get from monotonicity that
\begin{equation}
\label{dVdr2}
\forall~ D > 0~~\exists~ r_0~~\textrm{such that}~~\forall~ r > r_0~~~
r^2 \sqrt{W} > D
\end{equation}
and, since $g^{rr} < 1$, we also have $r^2 dV/dr > D$ for sufficiently large $r$. 
Integrating the latter relation between some $\widehat r$ and infinity we get a contradiction 
to (\ref{Vbd}). Therefore, $r^2 \sqrt{W}$ is bounded.

Using next the divergence theorem and $V_c \le V$ (where $V_c$ is the value of $V$ at the centre) we obtain

\begin{eqnarray}
\label{mass}
m(r) & = & \int_{0}^{r} \rho r'^2 dr' \le \int_{0}^{r}\rho \sqrt{g_{rr}} r'^2 dr'
\le V_c^{-1}\int_{0}^{r} V (\rho + 3p) \sqrt{g_{rr}}r'^2 dr' = {}
\nonumber\\
& & {} = V_c^{-1}\int_{r = const.} r^2 \sqrt{W} d\omega <
\infty.
\end{eqnarray}
Therefore, the ADM-mass 
$M  = \lim_{r \rightarrow \infty} m(r)$ exists, $g^{rr} \rightarrow 1$,
and the limit 

\begin{equation}
\label{limV}
\lim_{r \rightarrow \infty} r (1 - V)  
= \lim_{r \rightarrow \infty} r^2 \frac{dV}{dr} =  \lim_{r \rightarrow \infty} r^2 \sqrt{W} = N
\end{equation}
exists as well (and is called the "Komar mass").
 Moreover, we note that $\rho \in L^1_{0,-3}$ and, from (\ref{Eu}), that $p \in L^1_{1,-4}$.
Finally, using (\ref{DelV}) or (\ref{d2V}) again, we obtain
\begin{equation}
\label{dWdr}
\frac{dW}{dr} = - \frac{4}{r} W + 8 \pi V (\rho + 3p) \sqrt{g_{rr}W} \in L^1_{0,-5}.
\end{equation}

While these properties are slightly different from the AFMD conditions as defined in Sect.2., 
they still imply that the surface integral arising by applying
the divergence theorem to (\ref{DW}) vanishes. To see this we rewrite the
latter expression as follows (using spherical symmetry and multiplying with $r^2$),

\begin{equation}
\frac{d}{dr} \left[ \frac{r^2}{(1-V^2)^2}
\left( V^{-1} \sqrt{g^{rr}} \frac{dW}{dr} + \frac{8 W^{3/2}}{1-V^2} -
8\pi \sqrt{W} \left(\rho - \frac{1 - 5V}{1 - V}p \right) \right) \right],
\end{equation}
use the divergence theorem and insert (\ref{limV}), $r^4 W \rightarrow  N^2$,
 and (\ref{dWdr}). Recalling that $I_{1} \le 0$ implies $I_0 \le 0$,
the virial theorem (\ref{vir1}) now applies and we get finite extent. 

As remarked before, it also follows that $\widehat g$, $\widehat W$ and 
$\widehat W_0$ have $C^{1}$ extensions to the point at infinity $\Lambda$ of
a compactification $\widehat {\cal M} = {\cal M} \cup \Lambda$.
Hence (instead of using the virial theorem in the form (\ref{vir1})) we could 
also apply the maximum principle to (\ref{DelW}) to show finiteness.

$\hfill \Box$

\section{Discussion}

Here we discuss further the results on finiteness in the light of what we call the 
"quasipolytropic" EOS 
\begin{equation}
\label{qp}
p = K \rho^{\frac{n+1}{n}}\left[ 1 + f \left(\rho^{\frac{1}{n}}\right)
\rho^{\frac{1}{n}} \right]
\end{equation}
where  $K,n \in R^{+}$ and $f: [0,\lambda) \rightarrow R$ is a smooth
function. Obviously, for $f \equiv 0$, (but in general also for small
$\rho$), we are left with the polytropic EOS, while the limit $n = 0$ 
reduces to the incompressible case $\rho = const.$.
This quasipolytropic class seems to comprise all physically  
interesting EOS, in NT as well as in GR.

Let us first compare the polytropes in NT with  the "Generalized Buchdahl EOS" in GR 
obtained from (\ref{qp}) with the  choice $1/K = (n+1) \rho_{-}^{1/n}$ and 
$f(\rho^{1/n}) = (\rho_{-}^{1/n}  - \rho^{1/n})^{-1}$ for some constant $\rho_{-}$,
and for $\rho < \rho_{-}$, viz.

\parbox{5.5cm}
{\begin{eqnarray}
p = \frac{1}{n + 1} \rho_{-}^{-\frac{1}{n}}\rho^{\frac{n+1}{n}} \nonumber
\end{eqnarray}}
\hfill
\parbox{7.5cm}
{\begin{eqnarray}
\label{genBuch}
p = \frac{1}{n + 1} \rho^{\frac{n+1}{n}}
(\rho_{-}^{\frac{1}{n}}  - \rho^{\frac{1}{n}})^{-1}.
\end{eqnarray}}

For the Newtonian polytropes we now find that $I_0 = (n-5)p$ and 
$I_1 = \rho (n-5)/(n+1)$.
In GR, the integral
 (\ref{Gam}) for $\Gamma$ is still elementary but the expressions for $I_0$ and $I_1$ are
rather involved. Nevertheless, they still have the appealing property that the quantities 
$I_0$ and $I_1$ are negative iff $ n < 5$, zero for $n = 5$, 
(this is the "Buchdahl"- case considered above, c.f. (\ref{Bucheos}),(\ref{Buchsol})) 
and positive iff $n > 5$. Hence the (in)finiteness properties of all static solutions 
(satisfying also the other requirements of C.I and C.II) depend only on the EOS
and are obvious from the theorem. 

While in NT the polytropes are also (for special values of $n$) adiabates and 
therefore physically significant, the adiabates in GR  are neither polytropes nor 
do they coincide with (\ref{genBuch}). Instead, comparing again with the
Newtonian polytropes, they read \cite{MD}

\parbox{5.5cm}
{\begin{eqnarray}
\rho = \left( n + 1 \right)^{\frac{n}{n + 1}}
\rho_{-}^{\frac{1}{n + 1}} p^{\frac{n}{n + 1}} \nonumber
\end{eqnarray}}
\hfill
\parbox{7.5cm}
{\begin{eqnarray}
\label{ad}
\rho = \left( n + 1 \right)^{\frac{n}{n + 1}}
\rho_{-}^{\frac{1}{n + 1}} p^{\frac{n}{n + 1}} + np.
\end{eqnarray}}\\
The integral (\ref{Gam}) for $\Gamma$ is again elementary in the GR case of
(\ref{ad}). We get $I_0 \le 0$ (and hence finiteness for AFMD solutions) if  

\parbox{5.5cm}
{\begin{eqnarray}
{n \le 5} \nonumber
\end{eqnarray}}
\hfill
\parbox{7.5cm}
{\begin{eqnarray}
\label{plim}
p \le \frac{\rho_{-}}{n+1} \left( \frac{5}{n} - 1 \right)^{n + 1}.
\end{eqnarray}}\\
Thus adiabatic, asymptotically flat solutions whose pressure nowhere exceeds (\ref{plim}) 
are finite. The condition $I_1 \le 0$, which guarantees finiteness for spherically symmetric solutions, 
leads to a (smaller) upper bound on $p$ which can also be computed explicitly.

We now consider a fixed EOS in the general class (\ref{qp}) and the corresponding 
one-parameter family of static, spherically symmetric solutions. 
Rendall and Schmidt have shown \cite{RS} that,  
for $1 < n  < 5$ the fluid region is finite provided that the central density $\rho_c$ 
is below a critical value $\rho_{crit}$ , while for $n > 5$ the fluid is always infinite 
(with infinite  mass). More recently, Makino has proven that the fluid is always
finite for $1 < n < 3$. On the other hand, our theorem implies that finiteness for small 
$\rho_c$ is guaranteed in the range $0 < n < 5$, and allows us to estimate  $\rho_{crit}$.
Now the interesting remaining question is what happens for $0 < n < 5$ if the central
density is increased. For polytropes in GR, this question has been
investigated by Nilsson and Uggla \cite{NU} using a combination of dynamical systems techniques 
with numerical ones. They find that all solutions are finite for $n < 3.339$. For larger values they 
observe, somewhat surprisingly, a {\it discrete} set of central densities with infinite configurations, 
some of which have finite but others infinite mass. The mathematical side of these phenomena still 
deserves to be understood.

We finally remark that interesting results on finiteness have also been 
obtained for Vlasov-Poisson and Vlasov-Einstein systems by Rein and Rendall \cite{RR}.
\\ \\
{\it Acknowledgements.}
 I am indebted to J. M. Heinzle for carefully reading the draft, pointing out errors and suggesting 
improvements. I am also grateful to R. Beig, P. Bizo\v n, A. Rendall, R. Steinbauer and
C. Uggla for helpful discussions and correspondence.
This work was supported by Jubil\"aumsfonds der \"Osterreichischen
Nationalbank (Projekt 7942), and by Max-Planck-Institut f\"ur
Gravitationsphysik.

\end{document}